\documentstyle[aps,version2,preprint]{revtex}    
\begin{document}

\draft

\begin{title}
Electrons, pseudoparticles, and quasiparticles \\
in one-dimensional interacting electronic systems
\end{title}

\author{J. M. P. Carmelo$^{1}$, A. H. Castro Neto$^{2}$,
and N. M. R. Peres $^{1}$}
\begin{instit}
$^{1}$ Department of Physics, University of \'Evora,
Apartado 94, P-7001 \'Evora Codex, Portugal
\end{instit}
\begin{instit}
$^{2}$ Institute of Theoretical Physics, University of California,
Santa Barbara, CA 93106-4030
\end{instit}
\receipt{24 February 1995}

\begin{abstract}
We find the singular transformation between the electron
operator and the pseudoparticle operators for the Hubbard
chain. We generalize the concept of quasiparticle
to one-dimensional electronic systems which in 1D
refers to many-pseudoparticle objects. We obtain explicit
results for the electron renormalization factor, self energy,
and vertex functions at the Fermi points. We also
discuss the possible connection of our results to higher
dimensions and explore the possibilities of instabilities in
the interacting problem such as the formation of Cooper pairs.
\end{abstract}
\renewcommand{\baselinestretch}{1.656}   


\narrowtext

The unconventional electronic properties of novel materials such
as the superconducting coper oxides and synthetic
quasi-unidimensional conductors has attracted much attention to
the many-electron problem in spatial dimensions $1\leq$D$\leq 3$.
Although quantum liquids in dimensions  $1<$D$<3$ are, probably,
neither Fermi liquids (3D) nor Luttinger liquids (1D) but
have instead an intermediate physics, the complexity of
the problem requires a good understanding of {\it both} the
different and common properties of these two limiting cases.
While their different properties were the motivation for the
introduction of the concept of Luttinger liquid in 1D
\cite{Haldane}, the characterization of their common
properties is also of great interest because the latter are
expected to be present in dimensions $1<$D$<3$ as well. One
example is the Landau-liquid character common to Fermi
liquids and some Luttinger liquids which consists in the
generation of the low-energy excitations in terms of different
momentum-occupation configurations of quantum objects
(quasiparticles or pseudoparticles) whose forward-scattering
interactions determine the low-energy properties of the
quantum liquid. This generalized Landau-liquid theory was
introduced in Ref. \cite{Carmelo90}, which
refers to contact-interaction soluble problems
(shortly after the same kind of ideas were applied to
$1/r^2$-interaction integrable models \cite{Haldane91}).

The nature of interacting electronic quantum liquids in
dimensions  $1<$D$<3$, including the existence or non
existence of quasiparticles and Fermi surfaces, remains an
open question of crucial importance for the clarification of
the microscopic mechanisms behind the unconventional
properties of the novel materials.
Inspired by the results of Landau's theory of the Fermi
liquid in 3D, it is one of the aims of this Letter to introduce
the operator which creates a well defined elementary
excitation of the Fermi system, which we call
{\it quasiparticle}. This excitation is a transition
between two exact ground states of the interacting
electronic problem differing in the number of electrons
by one. When one electron is added to the electronic system
the number of these excitations {\it also} increases
by one. Naturally, its relation to the electron will depend on the
overlap between the states associated with this
and the quasiparticle and how close we are
in energy from the starting interacting ground state.
Therefore, in order to define the quasiparticle we
need to understand the properties of the actual ground
state of the problem as, for instance, is given
by its exact solution via the Bethe ansatz (BA).
The vanishing of the one-electron renormalization
factor in 1D does not necessarily implies the non
existence of the above quasiparticles and clarifying
the problem in the 1D limit is also important
for understanding how dimensionality changes
the physics of interacting electronic systems
in dimensions $1\leq$D$\leq 3$.

We consider here the Hubbard model
\cite{Lieb,Frahm,Carmelo92,Carmelo94}
in one dimension with a finite chemical potential $\mu$ and in the
presence of a magnetic field $H$,
\begin{eqnarray}
\hat{H} = -t\sum_{j,\sigma}\left[c_{j,\sigma}^{\dag }c_{j+1,\sigma}
+c_{j+1,\sigma}^{\dag }c_{j,\sigma}\right] +
U\sum_{j} [c_{j,\uparrow}^{\dag }c_{j,\uparrow} - 1/2]
[c_{j,\downarrow}^{\dag }c_{j,\downarrow} - 1/2]
- \mu \sum_{\sigma} \hat{N}_{\sigma } - 2\mu_0 H\hat{S}_z \, ,
\end{eqnarray}
where $c_{j,\sigma}^{\dag }$ and $c_{j,\sigma}$ are the creation and
annihilation operators, respectively, for electrons at the
site $j$ with spin projection $\sigma=\uparrow, \downarrow$.
In what follows $k_{F\sigma}=\pi n_{\sigma}$ and
$k_F=[k_{F\uparrow}+k_{F\downarrow}]/2=\pi n/2$, where
$n_{\sigma}=N_{\sigma}/N_a$ and $n=N/N_a$, and $N_{\sigma}$ and
$N_a$ are the number of $\sigma$ electrons and lattice sites,
respectively ($N=\sum_{\sigma}N_{\sigma}$). The results
refer to all finite values of $U$, electron densities
$0<n<1$, and spin densities $0<m<n$. This problem can be
diagonalized using the BA \cite{Lieb}.
This solution refers to a pseudoparticle operator basis, as
was well established in Refs. \cite{Carmelo92,Carmelo94}
(the pseudoparticle Hamiltonian parameters and phase shifts we
refer below are studied in detail in these papers). At constant values
of the electron numbers this description of the problem
is very similar to Fermi-liquid theory, except for two
main differences: (i) the $\uparrow $ and $\downarrow $ quasiparticles
are replaced by the $c$ and $s$ pseudoparticles and
(ii) the discrete pseudoparticle momentum (pseudomomentum) is
of the usual form $q_j={2\pi\over {N_a}}I_j^{\alpha}$
but the numbers $I_j^{\alpha}$ are not always integers.
They are integers or half-integers depending
whether the number of particles in the system is even or odd.
This plays a central role in the present problem.
The actual ground state of $(1)$ is described in terms of
the above $c$ and $s$ pseudoparticles which are created and
annihilated by {\it fermionic} operators $b^{\dag }_{q,\alpha}$
and $b_{q,\alpha}$, respectively (where $\alpha= c,s$).
The ground state, $|0;N_c = N_{\uparrow}+N_{\downarrow},
N_s = N_{\downarrow}\rangle$, and low-energy
Hamiltonian eigenstates can be completely
described in terms of the occupations of these excitations.
(Below we use often the alternative notation for the
ground state, $|0;N_{\sigma},N_{-\sigma}\rangle $.)
The $c$ and $s$ pseudoparticles are non-interacting at the
small-momentum and low-energy fixed point and the spectrum
is described in terms of bands in a pseudo-Brillouin zone which
goes between $q_c^{(-)}\approx -\pi$ and $q_c^{(+)}\approx \pi$
for the $c$ pseudoparticles and $q_s^{(-)}\approx -k_{F\uparrow}$
and $q_s^{(+)}\approx k_{F\uparrow}$ for the $s$
pseudoparticles. In the ground state these are occupied
for $q_{F\alpha}^{(-)}\leq q\leq q_{F\alpha}^{(+)}$,
where the pseudo-Fermi points are such that
$q_{Fc}^{(\pm)}\approx \pm 2k_F$ and
$q_{Fs}^{(\pm)}\approx \pm k_{F\downarrow}$.
Often, it is useful to introduce the quantum number
$\iota =\pm 1$ which defines the right ($\iota =1$) and
left ($\iota =-1$) pseudoparticles of momentum
$\tilde{q}=q-q_{F\alpha }^{(\pm )}$. At higher
energies and (or ) large momenta the pseudoparticles
start to interact (this is the price paid for the choice
of a fermionic statistics!) via zero-momentum transfer forward-scattering
processes. As in a Fermi liquid, these are associated
with $f$ functions whose values at the pseudo-Fermi
points define the Landau parameters,
$F_{\alpha\alpha'}^j={1\over {2\pi}}\sum_{\iota =\pm 1}(\iota )^j
f_{\alpha\alpha'}(q_{F\alpha}^{(\pm)},\iota q_{F\alpha '}^{(\pm)})$,
where $j=0,1$. Their expressions involve the pseudoparticle
group velocities $v_{\alpha }=\pm v_{\alpha }(q_{F\alpha}^{(\pm)})$
and the parameters $\xi^j_{\alpha\alpha '} =
\delta_{\alpha ,\alpha'}+\Phi_{\alpha\alpha '}(q_{F\alpha
}^{(+)},q_{F\alpha' }^{(+)})+(-1)^j\Phi_{\alpha\alpha
'}(q_{F\alpha }^{(+)},q_{F\alpha' }^{(-)})$, where
$j=0,1$ and $\Phi_{\alpha\alpha '}(q,q')$ is a two-pseudoparticle
phase shift.

Our task is finding the relationship between the electronic
operators $c^{\dag }_{k,\sigma}$ in momentum space and the
pseudoparticle operators $b^{\dag }_{q,\alpha}$. In this
Letter we solve the problem at the relative momenta of
ground-state pairs differing in $N_{\sigma }$ by one.
Notice that the electron excitation is {\it
not} an eigenstate of the interacting problem and, therefore,
when the electronic operator acts onto the ground state it
produces a multiparticle process in terms of the pseudoparticles.
The study of the above ground-state pairs of the interacting
problem $(1)$ reveals that their relative momentum equals
{\it precisely } the $U=0$ Fermi points, $\pm k_{F\sigma}$.
We consider the case when the electron has spin projection, say,
$\uparrow$ and momentum $k_{F\uparrow}$ (the construction is
the same for the case of spin projection $\downarrow$). We
define the quasiparticle operator, ${\tilde{c}}^{\dag
}_{k_{F\uparrow },\uparrow}$, which creates one quasiparticle
with spin projection $\uparrow$ and momentum $k_{F\uparrow}$ as,

\begin{eqnarray}
{\tilde{c}}^{\dag }_{k_{F\uparrow},\uparrow}
|0; N_c = N_{\uparrow} + N_{\downarrow},
N_s = N_{\downarrow} \rangle =
|0; N_c = N_{\uparrow}+ 1 + N_{\downarrow},
N_s = N_{\downarrow} \rangle.
\end{eqnarray}
The quasiparticle operator defines a one-to-one correspondence
between the addition of one electron to the system and the creation
of one quasiparticle, exactly as we expect from the Landau theory
in 3D: the electronic excitation,
$c^{\dag }_{k_{F\uparrow},\uparrow}
|0; N_c = N_{\uparrow} + N_{\downarrow},N_s = N_{\downarrow} \rangle$,
defined at the Fermi momentum but arbitrary energy, contains
a single quasiparticle, as we show below. We will study
this excitation as we take the energy to be zero, where the
problem is equivalent to Landau's.
Since we are discussing the problem of addition or removal of
one particle the boundary conditions play a crucial role.
When we add or remove one electron from the many-body system we
have to consider the transitions between states with integer and
half-integer quantum numbers $I_j^{\alpha}$. The transition between
two ground states differing in the number of electrons
by one is then associated with two different processes:
a backflow in the Hilbert space of the pseudoparticles with a shift
of all the pseudomomenta by $\pm\frac{\pi}{N_a}$ and
the creation of one or a pair of pseudoparticles at the pseudo-Fermi
points. We find that the backflow is described in terms
of a unitary operator,

\begin{eqnarray}
U_{\alpha}(\delta q) = \exp\left\{ - \delta q
\sum_{q'}[{\partial\over {\partial q'}}
b^{\dag }_{q',\alpha }]b_{q',\alpha }
\right\}.
\end{eqnarray}
 From the expressions of the ground-state pseudoparticle
generators the following relation between the quasiparticle and the
pseudoparticles follows,

\begin{equation}
\tilde{c}^{\dag }_{\pm k_{F\uparrow },\uparrow } =
b^{\dag }_{q_{Fc}^{(\pm)},c}
U^{\pm 1}_{s} \, ,
\hspace{1cm}
\tilde{c}^{\dag }_{\pm k_{F\downarrow },\downarrow } =
b^{\dag }_{q_{Fc}^{(\pm)},c}b^{\dag }_{q_{Fs}^{(\pm)},s}
U^{\pm 1}_{c } \, ,
\end{equation}
where $U^{\pm 1}_{\alpha }=U_{\alpha }
\left(\mp\frac{\pi}{N_a}\right)$.
According to Eq. $(4)$ the $\sigma $ quasiparticles are
many-pseudoparticle objects which recombine the
pseudoparticle colors $c$ and $s$ (charge and spin in the
limit $m = n_{\uparrow}-n_{\downarrow} \rightarrow 0$ \cite{Carmelo94})
giving rise to spin projection $\uparrow $ and $\downarrow $
and have Fermi surfaces at $\pm k_{F\sigma }$.
However, note that two-quasiparticle objects can be of
two-pseudoparticle character because the product of the
two corresponding many-pseudoparticle operators is such
that $U^{+ 1}_{\alpha }U^{- 1}_{\alpha }=\openone$, as for
the triplet pair $\tilde{c}^{\dag }_{+k_{F\uparrow },\uparrow }
\tilde{c}^{\dag }_{-k_{F\uparrow },\uparrow }=
b^{\dag }_{q_{Fc}^{(+)},c}b^{\dag }_{q_{Fc}^{(-)},c}$.

In order to relate the quasiparticle operators $\tilde{c}^{\dag
}_{\pm k_{F\sigma },\sigma }$ to the electronic operators
$c^{\dag }_{\pm k_{F\sigma },\sigma }$ we have combined
a generator pseudoparticle analysis and a suitable Lehmann
representation with conformal-field theory
\cite{Frahm,Carmelo94}. We measure the energy $\omega $
from the initial chemical potential $\mu (N_{\sigma},
N_{-\sigma})$ (ie, we consider $\mu (N_{\sigma},N_{-\sigma})=0$).
As in a Fermi liquid, we find that the
one-electron renormalization factor $Z_{\sigma}(\omega)$
has a crucial role in the above relation. This factor
is given by the small-$\omega $ leading-order term of
$|1-{\partial \hbox{Re}\Sigma_{\sigma}
(\pm k_{F\sigma},\omega)\over {\partial\omega}}|^{-1}$,
where $\Sigma_{\sigma} (k,\omega)$ is the $\sigma $ self
energy. Remarkably, we found the following low-$\omega $ expression,
$\hbox{Re}\Sigma_{\sigma} (\pm k_{F\sigma},\omega)=
\omega [1-{\omega^{-1-\varsigma_{\sigma}}\over
{a^{\sigma }_0+\sum_{j=1,2,3,...}a^{\sigma }_j\omega^{4j}}}]$,
where $a^{\sigma }_j$ are constants and
$\varsigma_{\uparrow }=-2+\sum_{\alpha }{1\over 2}
[(\xi^1_{\alpha c}-\xi^1_{\alpha s})^2+(\xi^0_{\alpha c})^2]$
and $\varsigma_{\downarrow }=-2+\sum_{\alpha }{1\over 2}
[(\xi^1_{\alpha s})^2 +(\xi^0_{\alpha c}+\xi^0_{\alpha s})^2]$
are $U$, $n$, and $m$ dependent exponents which
for $U>0$ are negative and such that
$-1<\varsigma_{\sigma}<-1/2$. Therefore, both the real part
of the Green function $\hbox{Re}G_{\sigma} (\pm k_{F\sigma},\omega)$
and the lifetime $\tau_{\sigma}=1/\hbox{Im}\Sigma_{\sigma}
(\pm k_{F\sigma},\omega)$ diverge when $\omega\rightarrow 0$ as
$\omega^{\varsigma_{\sigma}}$, yet $Z_{\sigma}(\omega)=
{a^{\sigma }_0\over{|\varsigma_{\sigma} |}}\omega^{1+
\varsigma_{\sigma}}$
vanishes in that limit and there is no overlap between the
quasiparticle and the electron, in contrast to a Fermi liquid.
In the different three limits $U\rightarrow 0$,
$m\rightarrow 0$, and $m\rightarrow n$ the
exponents $\varsigma_{\uparrow}$ and
$\varsigma_{\downarrow}$ are are equal
and given by $-1$, $-2+{1\over 2}[{\xi_0\over 2}+{1\over
{\xi_0}}]^2$, and $-{1\over 2}-\eta_0[1-{\eta_0\over 2}]$,
respectively. Here the $m\rightarrow 0$ parameter
$\xi_0$ changes from $\xi_0=\sqrt{2}$ at
$U=0$ to $\xi_0=1$ as $U\rightarrow\infty$
and $\eta_0=({2\over {\pi}})\tan^{-1}\left({4t\sin
(\pi n)\over U}\right)$.

Our method allows the identification of which particular
Hamiltonian eigenstates contribute to the above
$a^{\sigma }_j\omega^{4j}$ corrections and leads to the
following relation between the electron and quasiparticle
operators

\begin{equation}
c^{\dag }_{\pm k_{F\sigma},\sigma } =
\sqrt{Z_{\sigma }(\omega )|\varsigma_{\sigma}|}[1 +
\omega^2\sum_{\alpha ,\alpha ',\iota =\pm 1}
C_{\alpha ,\alpha '}^{\iota}
{\hat{\rho}}_{\alpha ,\iota } (\iota{2\pi\over {N_a}})
{\hat{\rho}}_{\alpha ',-\iota } (-\iota{2\pi\over {N_a}})
+ {\cal O}(\omega^4)]
\tilde{c}^{\dag }_{\pm k_{F\sigma},\sigma } \, ,
\end{equation}
where $C_{\alpha ,\alpha '}^{\iota}$ are constants
and ${\hat{\rho}}_{\alpha ,\iota } (k)=\sum_{\tilde{q}}
b^{\dag}_{\tilde{q}+k,\alpha ,\iota}
b_{\tilde{q},\alpha ,\iota}$ is a pseudoparticle-pseudohole
operator. When $\omega\rightarrow 0$ this relation
refers to a singular transformation because
$Z_{\sigma }(\omega )$ vanishes in that limit.
Combining Eqs. $(4)$ and $(5)$ gives the electron
operator in the pseudoparticle basis. The singular
nature of the transformation $(5)$ which maps
the 0-renormalization-factor electron
onto the 1-renormalization-factor quasiparticle
explains the perturbative character of the
pseudoparticle-operator basis \cite{Carmelo94}.
It is this perturbative character that determines
the form of expression $(5)$ which except for
the non-classical exponent in the $\sqrt{Z_{\sigma }(\omega )}
=\sqrt{{a^{\sigma}_0\over {|\varsigma_{\sigma}|}}}
\omega^{1+\varsigma_{\sigma}\over 2}$
factor (absorbed by the electron-quasiparticle transformation)
includes only classical exponents, as in a Fermi liquid.
Combining the relation $\tilde{c}^{\dag }_{k_{F\sigma },\sigma }
|0;N_{\sigma}, N_{-\sigma}\rangle = |0;N_{\sigma}+1,
N_{-\sigma}\rangle$ with Eq. $(5)$ we find that
$\sqrt{Z_{\sigma }(\omega )|\varsigma_{\sigma}|}=
|\langle 0;N_{\sigma}+1,N_{-\sigma}|c^{\dag }_{k_{F\sigma },\sigma }
|0;N_{\sigma},N_{-\sigma}\rangle|
\propto \omega^{1+\varsigma_{\sigma}\over 2}$. In
the present thermodynamic limit this result
is only valid in the limit $\omega\rightarrow 0$ and
confirms that this amplitude vanishes, the way it goes
to zero with the excitation energy being relevant for
comparison with other vanishing matrix elements:
the higher-order contributions to expression $(5)$
are associated with low-energy excited Hamiltonian
eigenstates orthogonal to the ground states and
whose matrix-element amplitudes
vanish as $\omega^{1+\varsigma_{\sigma}+4j\over 2}$
(with $2j$ the number of pseudoparticle-pseudohole
processes relative to $|0;N_{\sigma}+1,N_{-\sigma}\rangle $
and $j=1,2,3,...$). Therefore, the leading-order term
of $(5)$ and the exponent $\varsigma_{\sigma}$
fully control the low-energy overlap between the $\pm k_{F\sigma}$
quasiparticles and electrons and determines the
expressions of all $k=\pm k_{F\sigma }$ one-electron
low-energy quantities. For instance, we find for the
$\sigma $ spectral function $A_{\sigma}(\pm k_{F\sigma},\omega)
\propto \omega^{\varsigma_{\sigma}}$. (For a numerical
study see Ref. \cite{Muramatsu}.) Furthermore,
$A_{\sigma}(k,\omega)$ (and $\hbox{Re}G_{\sigma} (k,\omega)$)
vanishes when $\omega\rightarrow 0$ for all momentum values {\it except}
at the non-interacting Fermi-points $k=\pm k_{F\sigma}$.
It follows that for
$\omega\rightarrow 0$ the density of states,
$D_{\sigma} (\omega)=\sum_{k}A_{\sigma}(k,\omega)$,
results, exclusively, of contributions from
the peaks centered at $k=\pm k_{F\sigma}$
and is such that $D_{\sigma} (\omega)\propto
\omega A_{\sigma}(\pm k_{F\sigma},\omega)$.
It is known that $D_{\sigma} (\omega)\propto\omega^{\nu_{\sigma}}$,
where $\nu_{\sigma}$ is the exponent of
the equal-time momentum distribution expression,
$N_{\sigma}(k)\propto |k\mp k_{F\sigma}|^{\nu_{\sigma }}$
\cite{Lieb}. We find the relation $\varsigma_{\sigma}=\nu_{\sigma }-1$,
in agreement with the above analysis. However, this simple relation
does not imply that the equal-time expressions
provide full information on the small-energy instabilities.
For instance, in addition to the momentum values
$k=\pm k_{F\sigma}$ and in contrast to the spectral function,
$N_{\sigma}(k)$ shows singularities at
$k=\pm [k_{F\sigma}+2k_{F-\sigma}]$ \cite{Frahm}.
Therefore, only the direct low-energy study reveals all the
true instabilities of the quantum liquid.
(In some Luttinger liquids $N(k)\propto |k\mp k_F|^{\nu }$
with $\nu >1$ \cite{Solyom}. Then, $A(\pm k_F,\omega)
\propto\omega^{\nu -1}$ does not diverge.) The electron -
quasiparticle low-energy overlap also determines the behavior
of the two-electron vertex function at the Fermi momenta and
small energy. As usually \cite{Solyom}, we choose
the energy variables in such a way that
the combinations $\omega_1+\omega_2$,
$\omega_1-\omega_3$, and $\omega_1-\omega_4$ are all equal to
$\omega $. We find that the vertex function diverges as
$\Gamma_{\sigma\sigma '}^{\iota}(k_{F\sigma },\iota
k_{F\sigma '};\omega)\propto
\omega^{-2-\varsigma_{\sigma }-\varsigma_{\sigma '}}$,
where $\iota =\pm 1$.
Further, we could evaluate the following closed-form
expression $\Gamma_{\sigma\sigma '}^{\iota}(k_{F\sigma },\iota
k_{F\sigma '};\omega) = {1\over
{|\varsigma_{\sigma}\varsigma_{\sigma '}|
Z_{\sigma}(\omega)Z_{\sigma '}(\omega)}}
\{\sum_{\iota '=\pm 1}(\iota ')^{{1-\iota\over 2}}
[v_{\rho}^{\iota '} + (\delta_{\sigma ,\sigma '}
- \delta_{\sigma ,-\sigma '})v_{\sigma_z}^{\iota '}]
- \delta_{\sigma ,\sigma '}{\tilde{v}}_{\sigma }\}$,
where $v_{\rho}^{\iota }$ and $v_{\sigma_z}^{\iota }$
are given in Table I and involve only the
pseudoparticle velocities and Landau parameters
referred above. (We have not derived
the expression for the velocity ${\tilde{v}}_{\sigma }$.
Note, however, that the relevant quantity for the low-energy physics is
${\tilde{v}}_{\sigma }$-independent and reads
$\delta_{\sigma ,\sigma '}{\tilde{v}}_{\sigma }+
|\varsigma_{\sigma}\varsigma_{\sigma '}|
Z_{\sigma}(\omega)Z_{\sigma '}(\omega)
\Gamma_{\sigma\sigma '}^{\iota}(k_{F\sigma
},\iota k_{F\sigma '};\omega )$.)

Let us now consider the excitation energies
$\Delta E_{sp}=\omega^0_{\sigma ,-\sigma}
-\omega^0_{\sigma }-\omega^0_{-\sigma}$ and
$\Delta E_{tp\sigma }=\omega^0_{\sigma ,\sigma}
-2\omega^0_{\sigma }$, where
$\omega^0_{\sigma }=E_0(N_{\sigma }+1,N_{-\sigma })
-E_0(N_{\sigma },N_{-\sigma })$,
$\omega^0_{\sigma ,-\sigma}=E_0(N_{\sigma}+1,N_{-\sigma}+1)
-E_0(N_{\sigma},N_{-\sigma})$, and
$\omega^0_{\sigma ,\sigma}=E_0(N_{\sigma}+2,N_{-\sigma})
-E_0(N_{\sigma},N_{-\sigma})$ are ground-state
excitation energies. We have evaluated the
exact expressions of these vanishing energies which
to first order in ${1\over {N_a}}$ read,
$\Delta E_{sp}={\pi\over {N_a}}[v_c+F^0_{cc}+F^0_{cs}
-(v_s+F^1_{ss})+F^1_{cs}]$,
$\Delta E_{tp\uparrow }={\pi\over {N_a}}[v_c+F^0_{cc}
-(v_c+F^1_{cc})-(v_s+F^1_{ss})+2F^1_{cs}]$,
and $\Delta E_{tp\downarrow }={\pi\over {N_a}}[v_c+F^0_{cc}
+v_s+F^0_{ss}+2F^0_{cs}-(v_s+F^1_{ss})]$.
Again, these expressions involve exclusively
the velocities and Landau parameters.
Defining the hole concentration $\delta =1-n$,
while the above ground-state triplet pairing energies
are always positive, the singlet pairing energy
$\Delta E_{sp}$ is also positive except in a small concentration
domain, $0<\delta <\delta_c$, where it is negative.
This domain is larger for zero magnetization $m=0$.
In this case the $U$-dependent critical concentration
$\delta_c$ vanishes both in the limits $U\rightarrow 0$ and
$U\rightarrow \infty$ and is maximum for an intermediate but
relative large value of $U$. As in a Fermi liquid, we
find $\tilde{c}^{\dag }_{k_{F\sigma },\sigma }
\tilde{c}^{\dag }_{k_{F-\sigma },-\sigma }
|0;N_{\sigma},N_{-\sigma}\rangle =
|0;N_{\sigma}+1,N_{-\sigma}+1\rangle$ and
$\tilde{c}^{\dag }_{k_{F\sigma },\sigma }
\tilde{c}^{\dag }_{-k_{F\sigma },\sigma }
|0;N_{\sigma},N_{-\sigma}\rangle =
|0;N_{\sigma}+2,N_{-\sigma}\rangle$.
This implies that $\Delta E_{sp}$ and $\Delta E_{tp\sigma }$
are quasiparticle pairing energies. Therefore, for
$0<\delta <\delta_c$ there is attraction between
quasiparticles in the singlet Cooper-pair channel.
In a Fermi liquid this would imply a
singlet-superconductivity instability for hole concentrations
$0<\delta <\delta_c$. However, since in the limit
of vanishing energy the electrons and quasiparticles have
no overlap the quasiparticle attraction does
not necessarily imply the occurrence of such superconductivity
instability. We evaluated the corresponding response
functions with the results $Re \chi_{sc} (\pm 2k_F,\omega)$ and
$Im \chi_{sc} (\pm 2k_F,\omega)\propto
\omega^{\varsigma_{sc}}$, and
$Re \chi_{tc\sigma} (0,\omega)$ and
$Im \chi_{tc\sigma} (0,\omega)\propto
\omega^{\varsigma_{tp\sigma}}$, where
for $U>0$ the exponents are positive
and such that $0<\varsigma_{sc}<1$,
$0<\varsigma_{tc\uparrow}<1$, and
$0<\varsigma_{tc\downarrow}<2$. Therefore, in the 1D
Hubbard model the low-energy electron - quasiparticle
overlap is not strong enough for the quasiparticle -
quasiparticle attraction in the singlet channel
giving rise to a superconductivity instability.

One of the goals of this Letter was, in spite
of the differences between the Luttinger-liquid
Hubbard chain and 3D Fermi liquids,
detecting common features in these two limiting
problems which we expect to be present in electronic
quantum liquids in spatial dimensions
$1<$D$<3$. As in 3D Fermi liquids, we find that
there are Fermi-surface quasiparticles in
the Hubbard chain which connect ground states
differing in the number of electrons by one and
whose low-energy overlap with electrons
determines the one-electron $\omega\rightarrow 0$
divergences. For instance, in spite of the vanishing
electron density of states and renormalization
factor, we find that the spectral function
vanishes at all momenta values {\it except} at
the Fermi surface where it diverges (as a Luttinger-liquid power law).
While low-energy excitations are
described by $c$ and $s$ pseudoparticle-pseudohole
excitations which determine the $c$ and $s$
separation \cite{Carmelo94}, the quasiparticles
describe only ground-state -- ground-state transitions
and recombine $c$ and $s$ (charge and spin in
the $m\rightarrow 0$ limit) giving rise to the
spin projections $\sigma $ (see Eq. $(4)$).
Importantly, we have written the electron operator
at the Fermi surface in the pseudoparticle basis.
The vanishing of the electron renormalization
factor implies a singular character for the transformation
$(5)$ which leads to the above
quasiparticles with renormalization factor 1.
Our exact results have confirmed that the electron renormalization
factor can be related to a single matrix-element
amplitude \cite{Anderson,Metzner}. Although in the
Luttinger-liquid case the form of the vertex function
at the Fermi surface depends on the way that the energies
go to zero \cite{Solyom}, our study reveals that the Landau
parameters which control the quasiparticle interaction
energies $\Delta E_{sp}$ and $\Delta E_{tp\sigma }$ are
determined by the {\it finite} renormalized vertex
$lim_{\omega\to 0}[Z_{\sigma}(\omega)Z_{\sigma '}(\omega)
\Gamma_{\sigma\sigma '}^{\iota}(k_{F\sigma },
\iota k_{F\sigma '};\omega)]$, as in a Fermi liquid.
This justifies the {\it finite} forward-scattering functions
$f_{\alpha\alpha '}(q,q')$ and the perturbative character
of the Hamiltonian $(1)$ in the pseudoparticle
basis \cite{Carmelo94}. Finally, from the existence of Fermi-surface
quasiparticles both in the 1D and 3D limits, our results
suggest their existence for quantum liquids in dimensions
$1<$D$<3$ and we predict the main role of increasing
dimensionality being the strengthening of the electron -
quasiparticle overlap \cite{Carmelo95}. For instance, if
this leads to a finite (yet small) vanishing-energy overlap,
the pre-existing 1D singlet quasiparticle pairing induced by
electronic correlations could lead to a real superconductivity
instability.

We thank D. K. Campbell, F. Guinea, and K. Maki
for illuminating discussions. This research was supported by
the NSF under the Grant No. PHY89-04035.


\newpage
\centerline{TABLE}
\vspace{0.10cm}

\begin{tabbing}
\sl \hspace{2cm} \= \sl $v^{\iota}_{\rho }$
\hspace{3cm} \= \sl $v^{\iota}_{\sigma_z }$\\
$\iota = -1$ \> $v_c + F^1_{cc}$ \>
$v_c + F^1_{cc} + 4(v_s + F^1_{ss} - F^1_{cs})$ \\
$\iota = 1$ \> $(v_s + F^0_{ss})/L^0$ \>
$(v_s + F^0_{ss} + 4[v_c + F^0_{cc} + F^0_{cs}])/L^0$ \\
\label{tableI}
\end{tabbing}
Table I - Expressions of the parameters $v^{\iota}_{\rho }$ and
$v^{\iota}_{\sigma_z }$ in terms of the velocities
$v_{\alpha}$ and Landau parameters $F^j_{\alpha\alpha '}$,
where $L^0=(v_c+F^0_{cc})(v_s+F^0_{ss})-(F^0_{cs})^2$.

\begin{references}
\bibitem[1]{Haldane}
	F. D. M. Haldane, J. Phys. C {\bf 14}, 2585 (1981).
\bibitem[2]{Carmelo90}
	J. Carmelo and A. A. Ovchinnikov, Carg\`ese lectures,
	unpublished (1990); J. Phys.: Condens. Matter
	{\bf 3}, 757 (1991).
\bibitem[3]{Haldane91}
	F. D. M. Haldane, Phys. Rev. Lett. {\bf 66}, 1529 (1991);
	E. R. Mucciolo, B. Shastry, B. D. Simons, and
	B. L. Altshuler, Phys. Rev. B {\bf 49}, 15 197 (1994).
\bibitem[4]{Lieb}
	Elliott H. Lieb and F. Y. Wu, Phys. Rev. Lett. {\bf 20},
	1445 (1968).
\bibitem[5]{Frahm}
	Holger Frahm and V. E. Korepin, Phys. Rev. B {\bf 42},
	10 553 (1990); {\it ibid.} {\bf 43}, 5653 (1991);
	Masao Ogata, Tadao Sugiyama, and Hiroyuki Shiba,
	{\it ibid.} {\bf 43}, 8401 (1991).
\bibitem[6]{Carmelo92}
	J. M. P. Carmelo, P. Horsch, and A. A. Ovchinnikov,
	Phys. Rev. B {\bf 45}, 7899 (1992); {\it ibid.}
	{\bf 46}, 14 728 (1992).
\bibitem[7]{Carmelo94}
	J. M. P. Carmelo, A. H. Castro Neto, and
	D. K. Campbell, Phys. Rev. Lett. {\bf 73},
	926 (1994); {\it ibid.}, Phys. Rev. B {\bf 50}, 3667
	(1994); {\it ibid.} 3683 (1994); J. M. P.
	Carmelo and N. M. R. Peres, {\it ibid.} {\bf 51},
	7481 (1995).
\bibitem[8]{Muramatsu}
	R. Preuss, A. Muramatsu, W. von der Linden,
	P. Dierterich, F. F. Assaad, and W. Hanke,
	Phys. Rev. Lett. {\bf 73}, 732 (1994).
\bibitem[9]{Anderson}
	P. W. Anderson, Phys. Rev. Lett. {\bf 64}, 1839 (1990).
\bibitem[10]{Metzner}
	Walter Metzner and Claudio Castellani,
	preprint (1994).
\bibitem[11]{Solyom}
	J. S\'olyom, Adv. Phys. {\bf 28}, 201
	(1979); J. Voit, Phys. Rev. B {\bf 47},
	6740 (1993); Walter Metzner and Carlo Di Castro,
	{\it ibid.} 16 107 (1993).
\bibitem[12]{Carmelo95}
	J. M. P. Carmelo, F. Guinea, P. Horsch,
	and K. Maki, preprint (1995).
\end{references}
\end{document}